\documentclass[twocolumn,showpacs,preprintnumbers,amsmath,amssymb]{revtex4}

\usepackage{graphicx}
\usepackage{dcolumn}
\usepackage{bm}

\begin{document}

\preprint{APS/123-QED}

\title{A Unified Model of $\alpha$-Helix/$\beta$-Sheet/Random-Coil Transition in Proteins}

\author{Liu Hong}
\email{hong-l04@mail.tsinghua.edu.cn}
\author{Jinzhi Lei}%
\affiliation{%
Zhou Pei-Yuan Center for Applied Mathematics, Tsinghua University,
Beijing, P.R. China, 100084\\
}%

\date{\today}

\begin{abstract}
The theory of transition between $\alpha$-helix, $\beta$-sheet and
random coil conformation of a protein is discussed through a simple
model, that includes both short and long-range interactions. Besides
the bonding parameter and helical initiation factor in Zimm-Bragg
model, three new parameters are introduced to describe beta
structure: the local constraint factor for a single residue to be
contained in a $\beta$-strand, the long-range bonding parameter that
accounts for the interaction between a pair of bonded
$\beta$-strands, and a correction factor for the initiation of a
$\beta$-sheet. Either increasing local constraint factor or
long-range bonding parameter can cause a transition from
$\alpha$-helix or random coil conformation to $\beta$-sheet
conformation. The sharpness of transition depends on the competition
between short and long-range interactions. Other effective factors,
such as the chain length and temperature, are also discussed. In
this model, the entropy due to different ways to group
$\beta$-strands into different $\beta$-sheets gives rise to
significant contribution to partition function, and makes major
differences between beta structure and helical structure.
\end{abstract}

\pacs{Valid PACS appear here}
\maketitle

\section{Introduction}

The formation of protein secondary structure has attracted great
interest in the past 50 years, due to their remarkably regular
spatial arrangement\cite{BT}. There are usually two distinct
interactions involved: namely short-range interaction between atoms
and groups which are near neighbors in sequence along the chain; and
long-range interactions involving pairs of units which are remote in
sequence but near in space\cite{Flory}. How these interactions
affect the formation of secondary structure, especially
$\beta$-sheet, is essential in the study of protein folding. On the
other hand, a protein sequence can display structure ambivalence and
interconverts between $\alpha$-helix and $\beta$-sheet
conformations\cite{Patel:07}. This phenomenon may concern about many
well-known diseases, such as Alzheimer's, Mad Cow and Parkinson's
disease. But the underlying mechanism is still not very clear. In
this paper, we try to establish a unified statistical model of
$\alpha$-helix/$\beta$-sheet/random-coil transition in proteins,
which includes both short-range and long-range interactions. It will
be an extension of Zimm-Bragg's (Z-B) model to include beta
structures as well\cite{ZB}, and is totally different from former
tension-induced $\alpha$/$\beta$ transition
theory\cite{Birstein,David1,David2,Schor,Finkelstein}, which has
only mentioned short-range interactions.

$\alpha$-helix and $\beta$-sheet are most important secondary
structure elements in protein. An $\alpha$-helix is found when a
stretch of consecutive residues all have torsion angle pair $(\phi,
\psi)$ approximately $-60^o$ and $-50^o$, the allowed region in the
bottom left quadrant of the Ramachandran plot. In $\alpha$-helix,
the CO group of residue $n$ is bonded to the NH group of residue
$n+4$. Thus, it contains only short-range interactions. The
$\beta$-sheet, however, is built up of several segments of the
polypeptide chain. And long-range interactions are involved. These
segments, $\beta$-strands, are usually of $5\sim10$ residues long,
with $(\phi, \psi)$ angles in the broad structurally allowed region
in the upper left quadrant of the Ramachandran plot. The
$\beta$-strands are aligned adjacent to each other, such that
hydrogen bonds can form between CO groups of one strand and NH group
of an adjacent strand, and vice versa.

In the presence of long-range interactions, it is not sufficient to
distinguish states of a protein by only identifying the states of
each amino acid. Various bonding patterns of $\beta$-strands can
result in different states. Therefore, the different ways of
arranging $\beta$-strands into $\beta$-sheets have to be considered.
In this paper, we will establish a model, in which two sets of codes
are introduced to represent respectively the state of each residue
and connecting pattern of the $\beta$-strands. Accordingly, we are
able to write down the partition function, through which the
transition between alpha, beta and random coil structures of a
protein is discussed in detail.

When no $\alpha$-helix is presented, our model shows that a
coil-to-sheet transition occurs as the local constraint factor is
increased. And the sharpness of transition is determined by the
long-range bonding parameter and the initiation factor. The smaller
these two parameters are, the sharper the transition will be. This
result has similar feature as that in coil-to-helix transition,
which has been studied extensively by Z-B model. With the increment
of the local constraint factor, the normalized number of
$\beta$-strands increases firstly, and then decreases after passing
the transition point, due to the competition between short-range and
long-range interactions.

Helix/sheet transition is of great interest in this paper. In the
presence of long-range interactions, the entropy contribution to
partition function becomes more important. This is due to the ways
of connecting $\beta$-strands into $\beta$-sheets can be various,
even when the state of each residue is known. Unlike enthalpy
controlled helix/coil transition in Z-B model, helix/sheet
transition is driven by entropy. A transition from $\alpha$-helix to
$\beta$-sheet occurs, when either the local constraint factor of
beta residues or the long-range bonding parameter of $\beta$-stands
is increased. The sharpness of transition depends on the competition
between short-range and long-range interactions. Moreover, the
compromise between short-range and long-range interactions is able
to give rise to mixed $\alpha/\beta$ structure.

The length and temperature dependence of the transition are also
studied. Statistical study of protein structures reported in Protein
Data Bank (PDB) shows that the fractions of alpha and beta residues
are nearly unchanged when the chain is sufficiently long($n\geq50$).
Comparing our theoretical result with the data, we are able to
identify the values of parameters in our model. Interestingly, the
parameter values are close to the transition point. This indicates
that protein sequences in living cells are well selected by
evolution, to compromise the required properties such as stability,
flexibility and diversity. The temperature dependence is studied
based on above parameter values. Our results show that at low
temperature, $\alpha$-helix is dominated because of strong local
interactions; while at high temperature, beta structure is more
favored due to larger entropy contribution.

This paper will begin with an extension of Z-B model in Section
\ref{sec:2}, where connection properties of $\beta$-strands are
introduced to obtain a complete description for the state of a
protein. The partition function is obtained in Section \ref{sec:3},
by simple assumptions on the states of amino acid residues and
$\beta$-strands. In Section \ref{sec:4}, we discuss the transition
from random coil and $\alpha$-helix to $\beta$-sheet under different
conditions. At last, a brief conclusion is given in Section
\ref{sec:5}.

\section{Model}
\label{sec:2}

This section presents a model for the polypeptide chain that is
intended to extend Z-B model to include beta structures.
Specifically, it establishes a second order coding for the
connection of $\beta$-strands in $\beta$-sheets. The partition
function is then formulated as contributions from all $\alpha$, all
$\beta$ and $\alpha/\beta$ mixed structures. To describe the model
in detail, we firstly have to make a complete description of the
conformation of a chain, which will be finished in two steps.

The first step is to define the state of each amino acid residue, in
other words, which one is $\alpha$-helical and which one is in
$\beta$-strand. Definition of $\alpha$-helical residue is
straightforward, i.e., those residues whose NH group is bonded to
the CO group of the fourth preceding residue. Here we have assumed
that bonding of a residue, if it occurs in $\alpha$-helix, is always
to the fourth preceding residue, and disregard other helical
structures such as $\pi$-helix or $3_{10}$-helix\cite{Rohl}. To
determine whether a residue is in $\beta$-strand is sticky and
depends on the state of its neighboring residues. We assume that a
residue is regarded as in $\beta$-strand if the angle pair
$(\phi,\psi)$ of itself or both of its neighboring residues take
values in the upper left quadrant of Ramachandran plot. Now, a chain
of $n$ residue can be described by a sequence of $n$ symbols, each
of which can have one of three values: digit $0$ represents a random
coil residue, $1$ for an $\alpha$-helical residue, and $2$ for a
residue in $\beta$-strand. An example is shown by the 1st code in
Fig. \ref{fig:mod}.

Knowing the state of each residue is not sufficient. We need a 2nd
code to describe how $\beta$-strands are connected into
$\beta$-sheets. We assign a symbol $S_i^j$ for the $j$'th strand in
the $i$'th $\beta$-sheet(2nd code in Fig. \ref{fig:mod}). Here, the
number $j$ is assigned not according to the sequence along the
chain, but the  spatial connection position in the $\beta$-sheet.
Now, the state of a chain can completely be described by these two
sequences of codes as shown in Fig. \ref{fig:mod}. We will see later
that this second step makes major differences between $\beta$-sheet
and $\alpha$-helix, and gives rise to the helix/sheet transition.

\begin{figure*}[bt]
\centering
\begin{tabular}{lcccccccccccccccccccccccccccccccccccccccccccccccccccccccccccccccccccccccccc}
Peptide units: &1&2&3&4&5&6&7&8&9&$\cdot$&$\cdot$&$\cdot$\\
1st code: &0&0&0&0&0&1&1&1&1&0&0&0&0&2&2&2&2&2&2&0&0&0&0&0&2&2&2&2&2&0&0&0&0&0&2&2&2&2&2&0&0&$\cdots$\\
Weights: & 1&1&1&1&1&$\sigma s$&$s$&$s$&$s$&1&1&1&1&$w$&$w$&$w$&$w$&$w$&$w$&1&1&1&1&1&$w$&$w$&$w$&$w$&$w$&1&1&1&1&1&$w$&$w$&$w$&$w$&$w$&1&1&$\cdots$\\
2nd code:& &&&&&&&&&&&&&\multicolumn{6}{c}{$S_1^1$}&&&&&&\multicolumn{5}{c}{$S_1^2$}&&&&&&\multicolumn{5}{c}{$S_2^1$}&&&$\cdots$\\
Weights:& &&&&&&&&&&&&&\multicolumn{6}{c}{$\eta p$}&&&&&&\multicolumn{5}{c}{$p$}&&&&&&\multicolumn{5}{c}{$\eta p$}&&&$\cdots$\\

\end{tabular}
\caption{Coding and weight of residues in a chain.} \label{fig:mod}
\end{figure*}


Finally, we introduce the statistical weight for a given state of a
chain as the product of following factors, according to the coding
established above:
\begin{enumerate}
\item[(1)] The quantity unity for every $0$ (coil residue).
\item[(2)] The quantity $s$ for every $1$ that follows a $1$ (helical residue).
\item[(3)] The quantity $\sigma s$ for every $1$ that follows $\mu$ or more 0's (initiation of a helix).
\item[(4)] The quantity $w$ for every $2$ (residue in a $\beta$-strand).
\item[(5)] The quantity $p$ for every $S_i^j$ with $j\geq 2$ ($\beta$-strand).
\item[(6)] The quantity $\eta p$ for every $S_i^1$ (initiation of a $\beta$-sheet).
\item[(7)] The quantity $0$ for one of the following cases:
\begin{enumerate}
\item[(a)] every $1$ that follows $2$ or a number of $0$ less than $\mu$;
\item[(b)] every $2$ that follows $1$ or a number of $0$ less than $\mu$;
\item[(c)] every single $2$ between two $0$.
\end{enumerate}
\end{enumerate}
The effect of assumption (7) is that the secondary structures are
separated from each other by at least $\mu$ coil residues; and each
$\beta$-strand contains at least $2$ residues. The number $\mu$ is
varied from case to case. For example, two helices are separated by
at least $4$ residues, while two $\beta$-strands can be separated by
only two residues ($\beta$-turn). In this paper, we will take a
uniform value, say $\mu = 3$, for simplicity. We will see later that
the value of $\mu$ has little effect on the transition, especially
for long chains.

The meaning of the statistical weights are as follows. The first
three weights are the same as those in Z-B model\cite{Qian,Poland}.
The factor unity is arbitrarily assigned to coil residues, since
only the relative ratio is effective. The factor $s$ measures the
contribution of a helical residue, relative to a coil residue, to
the partition function. It contains a decrease due to restriction of
torsion angle and an enhancement because of the hydrogen bonding.
The factor $\sigma$ represents the decrease of weight $s$ for the
first unit in a helix, since the formation of first hydrogen bond
causes restriction of the freedom of residues between the bonded
ones.

The next three weights are new and for beta structures. The factor
$w$ measures the contribution of a residue in $\beta$-strand,
relative to a coil residue, to the partition function. It contains
an slight increase due to the lower energy compared to coil
residues\cite{Flory}. Note the torsion angle of beta residues can
take values from a quite broad region from the upper left quadrant
of the Ramachandran plot, thus the restriction in freedom is not
serious. The factor $p$ represents long-range interactions between a
pair of bonded $\beta$-strands. This includes a decrease due to
restriction of freedom of residues between the two bonded
$\beta$-strands, and an enhancement due to the hydrogen bonding.
Nevertheless, since two $\beta$-strands can be separated by long
distance along the chain, the effect of decreasing is dominant.
Therefore, the value of $p$ is usually small. Similar to the case of
helix, an initiation factor $\eta$ is introduced to measure the
decrease of weight $p$ for the first pair of $\beta$-strands in a
$\beta$-sheet. In summary, for the five weights in the model, the
factors $s$ and $w$ are usually sightly larger than unity, while
$\sigma$, $p$ and $\eta$ are all less than unity.

In the above discussion, we assume that the bonding energy(factor
$p$) of each pair of $\beta$-strands are the same, and independent
of the number of both residues and hydrogen bonds between the bonded
$\beta$-strands. It is a highly simplified representation of the
problem and enables us to write down the partition function. One may
introduce a set of factors $p(n_1, n_2)$, which depends on those two
values, to make the model more realistic. But since bonding patterns
of $\beta$-sheet varies greatly, our present knowledge is too
incomplete to justify a more refined model.

\section{Mathematical treatment}
\label{sec:3}

A formal representation of partition function $Z$ for a chain of $n$
residues can be obtained from above model by direct enumerating of
the number of different ways of arranging digits $0, 1, 2$ and the
state of strands $S_i^j$.

Let $X = (n_a, n_b, l_b, l_b, k_b)$ be the state of a chain, where
$n_a$, $n_b$, $l_a$, $l_b$, $k_b$ are respectively the number of
helical residues, beta residues, $\alpha$-helices, $\beta$-strands
and $\beta$-sheets. Then the statistical weight of state $X$ is
given by
\begin{equation}
 \label{eq:Q}
Q(X) = s^{n_a} w^{n_b} \sigma^{l_a} p^{l_b} \eta^{k_b}.
\end{equation}
Let $S(X)$ be the number of states, which has the same statistical
weight as $X$, and $\Omega$ be the phase space of possible states,
then the partition function is represented as
\begin{equation}
 \label{eq:Z}
Z = \sum_{X\in \Omega}Q(X)S(X).
\end{equation}
Explicitly, the phase space $\Omega$ is given by
$$\Omega = \left\{X\in \mathbb{Z}^5:\ \ \  \begin{array}{l}
0\leq k_b\leq l_b/2 \leq n_b/4 \leq \lfloor\frac{n}{4 + 2\mu}\rfloor,\\
0\leq l_a \leq n_a \leq \lfloor\frac{n}{1 + \mu}\rfloor, \\
n - (n_a + n_b) \geq \mu ( l_a + l_b)\\
l_b = n_b = 0\ \ \mathrm{if}\ k_b = 0\\
n_a = 0 \ \ \mathrm{if}\ l_a = 0
          \end{array}\right\}.$$
The degeneracy $S(X)$ is formulated as
\begin{equation}
 \label{eq:S}
\begin{array}{rcl}
S(X) &=& C_{n_a- 1}^{l_a - 1}C_{n_b-l_b - 1}^{l_b - 1}C_{l_b - k_b -1}^{k_b - 1} \\
&&\times  C_{n - n_a - n_b - (\mu-1) (l_a+l_b)}^{l_a + l_b}C_{l_a +
l_b}^{l_a} .
\end{array}
\end{equation}
This formula is obtained by following steps. Firstly, we partition
$n_a$ amino acids into $l_a$ $\alpha$-helices, $n_b$ amino acid into
$l_b$ $\beta$-strands and $l_b$ $\beta$-strands into $k_b$
$\beta$-sheet along the chain. Then we insert $n-n_a-n_b$ coil
residues between $\alpha$-helices and $\beta$-strands. Finally, we
rearrange the order of $\alpha$-helices and $\beta$-strands. The
meaning of each term is explained as follows. The binomials
$C_{n_a-1}^{l_a-1}$ represents the number of ways to group $n_a$
helical residues into $l_a$ helices. The terms
$C_{n_b-l_b-1}^{l_b-1}$ and $C_{l_b-k_b-1}^{k_b-1}$ are similar,
representing the number of ways to group beta residues into
$\beta$-strands and $\beta$-strands into $\beta$-sheets,
respectively. Here each $\beta$-sheet has at least two strands; and
each $\beta$-strand has at least two residues. The binomial $C_{n -
n_a - n_b - (\mu-1) (l_a+l_b)}^{l_a + l_b}$ represents how many ways
to insert $n-n_a-n_b$ coil residues between the secondary structure
elements, such that consecutive helix and strand are separated by at
least $\mu$ coil residues. The last factor $C_{l_a+l_b}^{l_b}$
represents how many ways to arrange the order of $\alpha$-helices
and $\beta$-strands. Note that here the $\alpha$-helices are
considered as identical. The $\beta$-strands are assumed to connect
to only the neighboring ones. Long-distance connections in
$\beta$-sheets are neglected, since the usual adopted arrangements
are quite limited at room temperature. One may propose other
assumptions. For example, $\beta$-strands are assumed to be able to
connect to each other freely ($l_b!C_{l_a+l_b}^{l_b}$), despite
their length and distance. Or all $\alpha$-helices and
$\beta$-strands are considered as distinguishable ($(l_a+l_b)!$), in
case of very strong long-range interactions between secondary
structure elements. Different assumptions may give different
results, but the general property of transition is almost unchanged,
according to our simulations. Further discussions are shown in the
appendix.

The partition function can be rewritten according to the structure
of the chain as follows:
\begin{equation}
 \label{eq:2}
Z = 1 + Z_{\alpha} + Z_{\beta} + Z_{\alpha/\beta}
\end{equation}
where
\begin{eqnarray}
\label{eq:za}
Z_{\alpha} &=& \sum_{l_a = 1}^{\lfloor \frac{n}{1+\mu}\rfloor} \sigma^{l_a} \sum_{n_a = l_a}^{n - \mu l_a} s^{n_a} C_{n_a-1}^{l_a-1} C_{n - n_a - (\mu-1) l_a}^{l_a}\\
 \label{eq:zb}
Z_{\beta} &=& \sum_{k_b = 1}^{\lfloor \frac{n}{4+2\mu} \rfloor } \eta^{k_b} \sum_{l_b = 2 k_b}^{\lfloor \frac{n}{2+\mu}\rfloor} p^{l_b}  C_{l_b - k_b - 1}^{k_b - 1}\nonumber\\
&&{}\times \sum_{n_b = 2l_b}^{n-\mu l_b} w^{n_b}  C_{n_b-l_b - 1}^{l_b - 1} C_{n - n_b - (\mu-1)l_b}^{l_b}\\
\label{eq:zab} Z_{\alpha/\beta} &=&\sum_{k_b=1}^{\lfloor
\frac{n}{4+2\mu}\rfloor}\eta^{k_b}\sum_{l_b=2k_b}^{\lfloor
\frac{n}{2+\mu}\rfloor}p^{l_b} C_{l_b-k_b-1}^{k_b-1}\nonumber\\
&&{}\times \sum_{n_b=2l_b}^{n-\mu
l_b}w^{n_b}C_{n_b-l_b-1}^{l_b-1}\sum_{l_a=1}^{\lfloor
\frac{n-n_b-\mu l_b}{1+\mu}\rfloor}
\sigma^{l_a}C_{l_a+l_b}^{l_a}\\
&&{}\times \sum_{n_a=l_a}^{n-n_b - \mu  ( l_a +l_b)}s^{n_a}
C_{n_a-1}^{l_a-1} C_{n-n_b-n_a - (\mu-1)
(l_a+l_b)}^{l_a+l_b}\nonumber
\end{eqnarray}
Here $Z_\alpha$ (or $Z_\beta$) represents the partition function for
the states of all $\alpha$ (or $\beta$) structures, and
$Z_{\alpha/\beta}$ for $\alpha/\beta$ mixed structures.

In the following, we will show how the $\alpha$-helix, $\beta$-sheet
and random coil conformation at equilibrium state transit to each
other, when the parameters are changing. We define the fraction of
helical and beta residues respectively as
\begin{equation}
\label{eq:a} \theta_a:=\frac{n_{\alpha}}{n} =
\frac{1}{n}\dfrac{\partial \ln Z}{\partial \ln s}, \;
\theta_b:=\frac{n_{\beta}}{n} = \frac{1}{n} \dfrac{\partial \ln
Z}{\partial \ln w}.
\end{equation}
The normalized number of $\alpha$-helices and $\beta$-strands are
also of interest.
\begin{equation}
\label{eq:b} \phi_a:=\frac{l_{\alpha}}{n} =
\frac{1}{n}\dfrac{\partial \ln Z}{\partial \ln \sigma}, \;
\phi_b:=\frac{l_{\beta}}{n} = \frac{1}{n} \dfrac{\partial \ln
Z}{\partial \ln p}.
\end{equation}

\section{Results and Discussions}
\label{sec:4}

From discussions in previous section, it is easy to show that in the
absence of long-range interaction, the partition
function($Z=1+Z_\alpha$) is the same as that for helix/coil
transition, which has been well studied by Z-B model. In following,
we will focus on how the parameters induce transitions between
$\alpha$-helix, $\beta$-sheet and random coil. From
\eqref{eq:za}-\eqref{eq:zab}, the parameter $\mu$ has only minor
effect on the results. Here, we will set  $\mu = 3$ for the minimum
number of coil residues between secondary structures.

\subsection{Sheet/Coil Transition}

At first, we study the transition from random coil to regular
$\beta$-sheet($Z=1+Z_\beta$), which is similar to helix/coil
transition\cite{ZB,Birshtein,Grosberg}.

\begin{figure}[h]
\includegraphics[width=0.5\textwidth,height=0.4\textwidth]{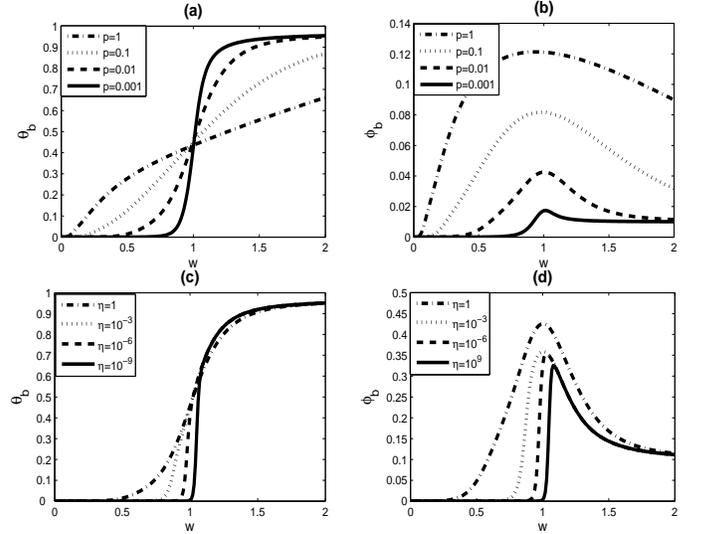}
\caption{\label{bct}~Sheet/coil transition ($n=200$).  The
dependence of $\theta_b$ (a) and $\phi_b$ (b) on $w$, with different
values of $p$, and $\eta$ is set to $1.0$.  The dependence of
$\theta_b$ (a) and $\phi_b$ (d) on $w$, with different values of
$\eta$, and $p$ is set to $0.01$.}
\end{figure}

In this case, We have three tuneable parameters: $w$ for local
constraint of single residue, $p$ for long-range bonding
interactions between a pair of bonded $\beta$-strands and $\eta$ for
initiation correction of a $\beta$-sheet. The dependence of
$\theta_b$ and $\phi_b$ on the parameters are shown at Fig.
\ref{bct}.

From Fig. \ref{bct}(a), \ref{bct}(c), we can see that when $w$ is
increasing, the fraction of beta residues $\theta_b$ goes from zero
to unity monotonously. This means the random coils turn into regular
$\beta$-sheets when the local interactions become stronger. This
transition usually happens at $w=1$, in accordance with Z-B model.
The sharpness of transition depends on $p$ and $\eta$. The smaller
$p$ or $\eta$ is, the sharper the transition will be.

Unlike the fraction of $\beta$ residues, which increases
monotonously with respect to $w$, the number of $\beta$-strands
increases firstly to reach a peak at around the transition point,
and then decreases when $w$ keeps increasing (Fig.
\ref{bct}(b),(d)). We argue that this is a consequence of the
competition between short and long-range interactions. Generally
speaking, when $w$ is less than unity, short-range interactions
prompt long-range interactions. With the increment of $w$, there
tends to be more $\beta$-strands, as well as the strands get longer.
Thus $\theta_b$ and $\phi_b$ increase simultaneously. When $w$ is
larger than unity, short-range interactions suppress long-range
interactions. As a consequence, when $w$ is increasing further,
$\theta_b$ is almost unchanged while $\phi_b$ drops continuously. It
means that there tends to be less $\beta$-strands and the strands
still grow longer and longer.

\subsection{Helix/Sheet Transition}

\begin{figure}[h]
\includegraphics[width=0.48\textwidth,height=0.6\textwidth]{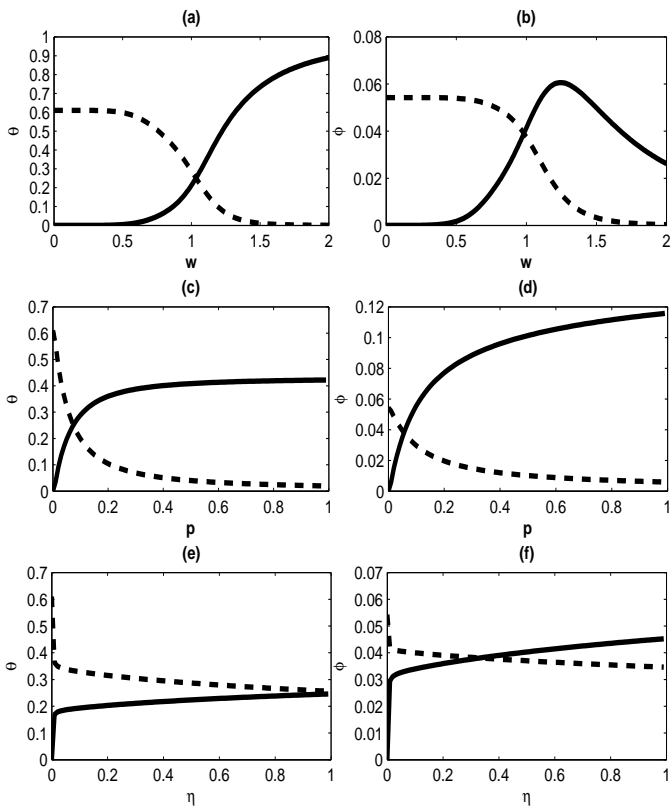}
\caption{\label{abt}~Helix/sheet transition ($n=200$).  In
(a)(c)(e), the dashed lines are for $\theta_a$; while the solid
lines are for $\theta_b$. In (b)(d)(f), the dashed lines are for
$\phi_a$; and the solid lines are for $\phi_b$. (a) The dependence
of $\theta_a$, $\theta_b$ on $w$. (b) The dependence of $\phi_a$,
$\phi_b$ on $w$. (c) The dependence of $\theta_a$, $\theta_b$ on
$p$. (d) The dependence of $\phi_a$, $\phi_b$ on $p$. (e) The
dependence of $\theta_a$, $\theta_b$ on $p$. (f) The dependence of
$\phi_a$, $\phi_b$ on $p$. In all computations, the chain length is
$n=200$. The parameters used are, except particular indicated,
$s=1.15$, $\sigma=0.037$, $w=1.03$, $p=0.072$ and $\eta=1$. }
\end{figure}

A more interesting result is $\alpha$-helix/$\beta$-sheet
transition. Since the major difference of $\beta$-sheet from
$\alpha$-helix is the existence of long-range interaction, which is
described by parameters $p$ and $\eta$. Thus not only the increase
of local effect $w$, but also the increment of long-range
interaction parameters $p$ and $\eta$ can induce a transition from
helical structure to beta structure(Fig. \ref{abt}).

From Fig. \ref{abt}(a,b), we can see that $\alpha$-helices transit
to $\beta$-sheets with the increment of short-range interaction $w$.
When $w$ is further increasing, the short-range interaction takes
over long-range interactions, and therefore the normalized number of
$\beta$-strands $\phi_b$ decreases.

From Fig. \ref{abt}(c)-(e), we can see that the fraction of beta
structure is very sensitive to the long-range interactions ($p$ and
$\eta$) when the parameters are small ($p, \eta\ll1$). This is a
consequence of large entropy and high cooperativity of beta
structures. Yet when the long-range interactions are strong enough
($p, \eta\sim1$), the dependence of $\theta$ and $\phi$ on the
parameters are far less evident.

In the present of long-range interactions, the entropy contribution
to the partition function become important. This is originated from
two sources: different methods of arranging secondary structure
segments along the chain; and the ways of connecting $\beta$-strands
into $\beta$-sheets, which is specified for beta structure. In
protein, helical structure tends to have lower energy due to local
hydrogen bonds, while the beta structure has larger entropy. As
consequence, coil/helix transition is enthalpy controlled, while
helix/sheet transition is entropy driven. Moreover, the longer the
chain is, the sharper the helix/sheet transition will be. This
helix/sheet transition picture may shed light on the mechanisms of
structure transition of prion\cite{Prusiner} and other
proteins\cite{Patel:07}.

\subsection{Length Dependence}

\begin{figure}[h]
\centering (a)\\
\includegraphics[width=0.45\textwidth,height=0.35\textwidth]{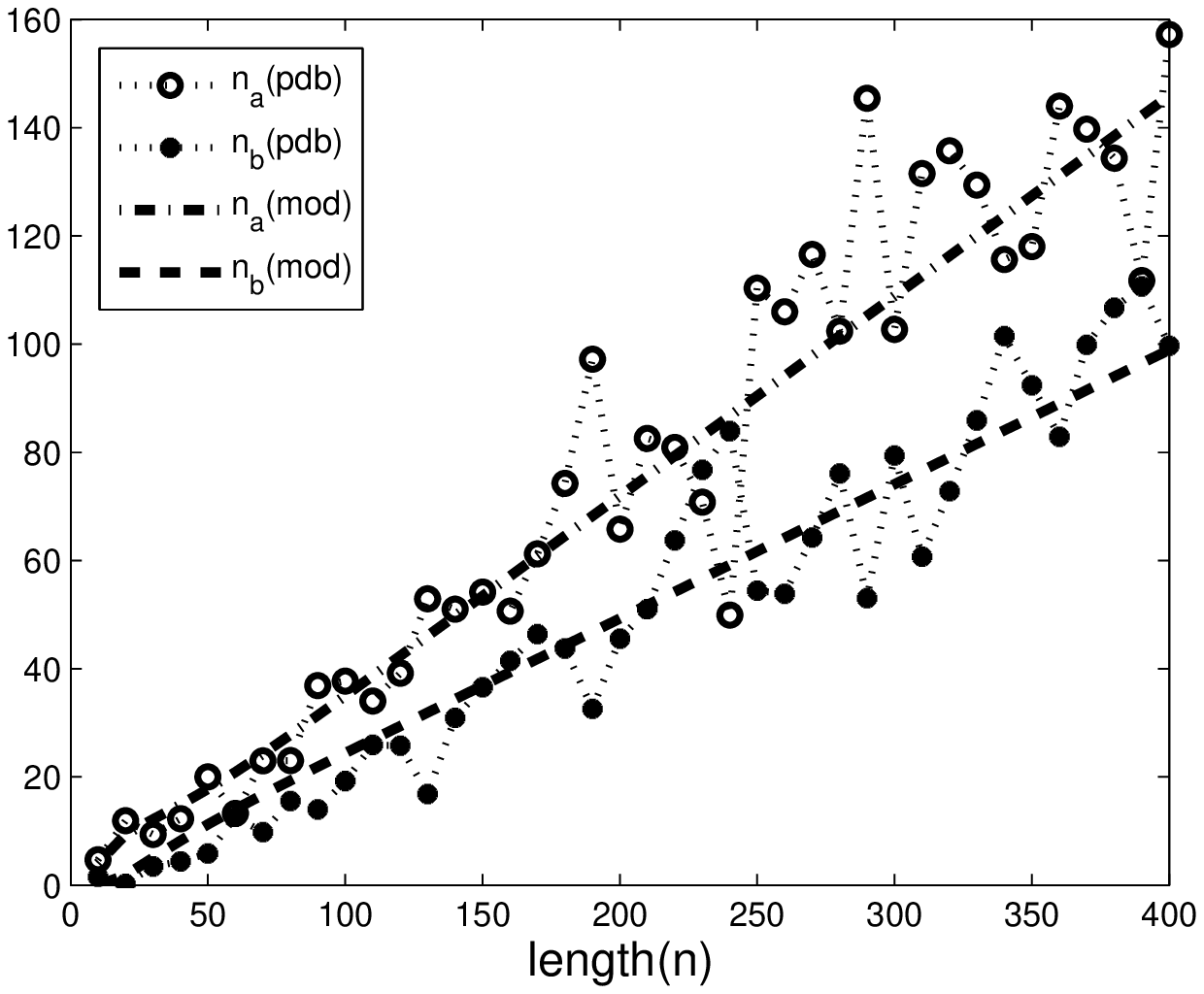}\\
(b)\\
\includegraphics[width=0.45\textwidth,height=0.35\textwidth]{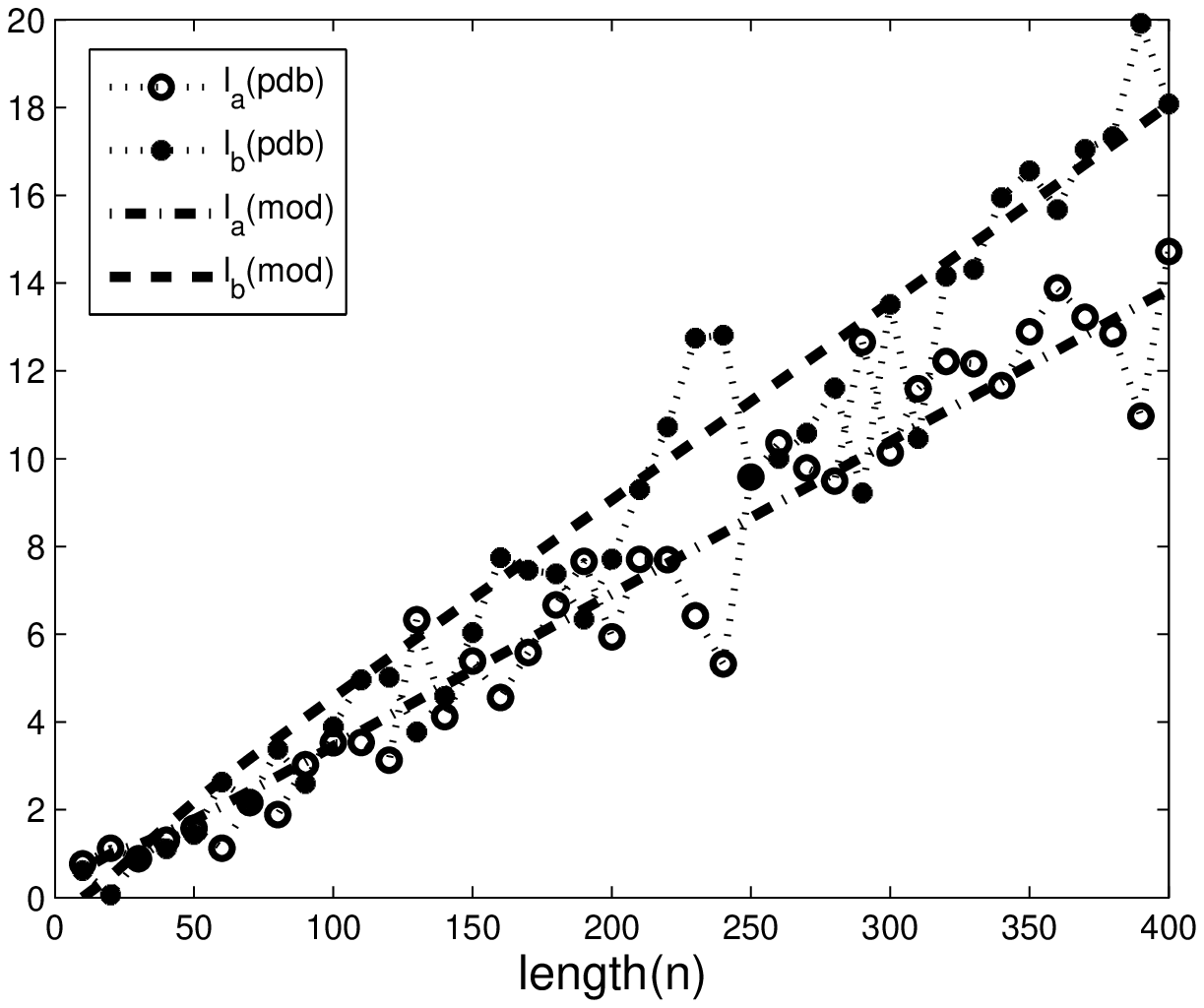}\\
\caption{\label{length}~Experimental data and theoretical simulation
for the length dependence. The data are fitted by the model with
parameters $w = 1.03, p = 0.072, \eta=1, s=1.15, \sigma=0.037$. (a)
The average number of alpha and beta residues. (b) The average
number of $\alpha$-helices and $\beta$-strands.}
\end{figure}

The length dependence of the average fractions of alpha and beta
residues obtained by studying the native structures in Protein Data
Bank is plotted in Fig. \ref{length}. The results show that despite
the variety from one protein to another, the average fractions are
roughly independent to the chain length, except for very short
chains ($n\leq 30$), with about $32\%$ alpha residues and $25\%$
beta residues.  The data can be fitted by our model with parameters
$w = 1.03, p = 0.072, \eta=1, s=1.15, \sigma=0.037$ (Fig.
\ref{length} (a)). Note that here the number of alpha residues from
our model is given by $n_{\alpha}=n (\theta_a+3\phi_a)$. Since in
our model, the first three residues in every helix is not treated as
helical. The same set of parameters also gives good agreement for
the average number of $\alpha$-helices and $\beta$-strands(Fig.
\ref{length}(b)). Accordingly, we obtain the average length of
$\alpha$-helix to be $11$, and $\beta$-strand to be $6$, both agree
with experimental data.

In previous simulation results(Fig. \ref{abt}), we can see that
while fitting our model to PDB data, the parameters take values
around the helix/sheet transition point. We argue that this is not
by chance, but is a prerequisite for proteins in living cells. Since
diverse structure of proteins in living cells is required for their
biological functions, particular values of the parameters are needed
to compromise the required properties such as stability, flexibility
and diversity, by natural selection.

\subsection{Temperature Dependence}

\begin{figure}[h]
\centering
\includegraphics[width=0.45\textwidth,height=0.35\textwidth]{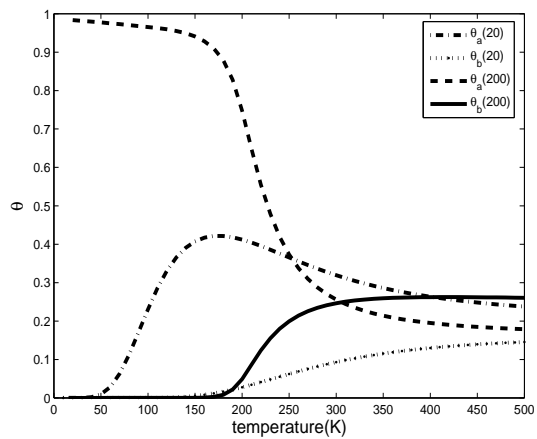}\\
\caption{\label{temp}~Temperature dependence of fractions of alpha
and beta resides for both short ($n=20$) and long ($n=200$) chains.
Here $T_0=300K, w = 1.03, p = 0.072, \eta=1, s=1.15, \sigma=0.037$.}
\end{figure}

From previous discussion, helical structure tends to have low
enthalpy, and beta structure tends to have large entropy. Thus,
under certain condition, we should have a transition from helical
structure to beta structure when the temperature is increased. To
study this temperature induced transition, we need to explore the
dependence of parameters on temperature. For the parameter $s$, for
instance, we have\cite{LR}
$$s = \exp (\Delta G/k_B T)$$
where $T$ is the temperature. Here $\Delta G$ is the free energy
change by converting one residue from random coil state to helical
state. Assume that $\Delta G$ is independent to the temperature.
Then the parameters $s_i$ for corresponding temperatures $T_i$
($i=0,1$) are related by
$$s_1 = s_0^r$$
where $r=T_0/T_1$ is the ratio of original and final temperature.
Let $T_0=300K$ be the room temperature, thus we can investigate the
temperature induced structure transition by changing the system
parameters simultaneously
$$(\sigma, s, \eta, w, p) \rightarrow (\sigma^r, s^r, \eta^r, w^r, p^r)$$
with $r = T/T_0$. The simulation results are shown in Fig.
\ref{temp}.

The transition from alpha to beta structure, when the temperature is
increasing, is obvious in our simulation. For long chains, at low
temperature, $\alpha$-helix is dominated because of strong local
interactions; while at high temperature, beta structure is more
favored for larger entropy contribution. But for short chains
($n<=26$ in this case), $\alpha$-helix becomes unstable at low
temperature, and $\theta_a$ drops to zero. A big drawback of the
results in Fig. \ref{temp} is that the regular secondary structures
will not break into random coils at high temperature, by which the
proteins should unfold. This is mainly due to our strong assumption
that $\Delta H$ does not change with the temperature, which is not
true when the temperature is far from the room temperature.

\section{Conclusion}
\label{sec:5}

In this paper, we extend the Z-B model to include beta structures,
by introducing two sets of codes which represent respectively the
state of each amino acid residue and connecting pattern of the
$\beta$-strands. In additional to the bonding parameter and helical
initiation factor in Z-B model, three new parameters are introduced
to describe beta structures: the local constraint factor for a
single residue to be contained in a $\beta$-strand, the long-range
bonding parameter that accounts for the interaction between a pair
of bonded $\beta$-strands, and a correction factor for the
initiation of a $\beta$-sheet. Then the partition function is
obtained based on our model, which contains both short and
long-range interactions.

Through numerical study, the transition from random coil and
$\alpha$-helical conformation to $\beta$-sheet structure are
discussed respectively. In common, either the increase of the
short-range or the long-range interactions of beta structure can
cause a transition. And the sharpness of transition is mainly
determined by the long-range bonding parameter and the initiation
factor. However, the coil/sheet transition is enthalpy controlled,
while helix/sheet transition is entropy driven. Other effective
factors, such as the chain length and temperature, are also shown.
The PDB data for protein chains with different length are fitted by
our model and show a fairly well agreement. In this way, we can
identify the values of parameters in our model. Interestingly, these
values are close to the transition point. This indicates that
protein sequences in living cells are well selected by evolution, to
compromise the required properties such as stability, flexibility
and diversity. At last, by considering the relationship between
statistical weights and free energy, a temperature induced
helix/sheet transition is observed. Our results show that at low
temperature, $\alpha$-helix is dominated because of strong local
interactions; while at high temperature, beta structure is more
favored due to large entropy contribution.

We hope our results may shed light on the mechanisms of structure
transition in prion and other proteins, as well as the understanding
of short and long-range interactions in the formation of secondary
structures.

\begin{acknowledgments}
We thank Professor Kerson Huang  and Professor C.C.Lin for their
many helpful discussions.
\end{acknowledgments}

\appendix

\section{}
\begin{figure}[h]
\centering
\includegraphics[width=0.45\textwidth,height=0.35\textwidth]{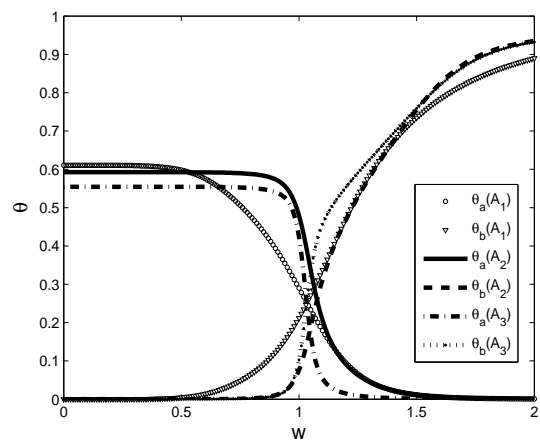}\\
\caption{\label{model1}~Comparison of helix/sheet transition under
three different assumptions. Case $A_1$ with $s=1.15$,
$\sigma=0.037$, $p=0.072$ and $\eta=1$. Case $A_2$ with $s=1.14$,
$\sigma=0.04$, $p=0.0062$ and $\eta=1$. Case $A_3$ with $s=1.14$,
$\sigma=0.0047$, $p=0.007$ and $\eta=1$. The values of parameters in
case $A_i(i=1,2,3)$ are the same as in Fig. \ref{model2},
respectively.}
\end{figure}

The exact factor for different ways of arranging $\alpha$-helices
and $\beta$-strands is undetermined in our present model, since our
current knowledge about the secondary structure arrangement in
natural proteins is too incomplete to justify the models. In
general, we have three extreme cases. If the connections are highly
restricted, i.e., $\beta$-strands are assumed to connect to only the
neighboring ones, the factor $C_{l_a+l_b}^{l_b}$ would be a good
estimate (Case $A_1$). On the other extreme, if the $\beta$-strands
can connect to each other freely, despite their long distance along
the chain, the factor should be $l_b!C_{l_a+l_b}^{l_b}$ (Case
$A_2$). But this can only happen at very high temperature. Moreover,
we can even assume that all $\alpha$-helices and $\beta$-strands are
distinguishable, in case of very strong long-range interactions
between secondary structure elements. Then the factor will be
$(l_a+l_b)!$ (Case $A_3$). Different assumption may give different
results, but the general properties of transition are almost
unchanged, which is shown in Fig. \ref{model1}.

\begin{figure}[h]
\centering
\includegraphics[width=0.45\textwidth,height=0.35\textwidth]{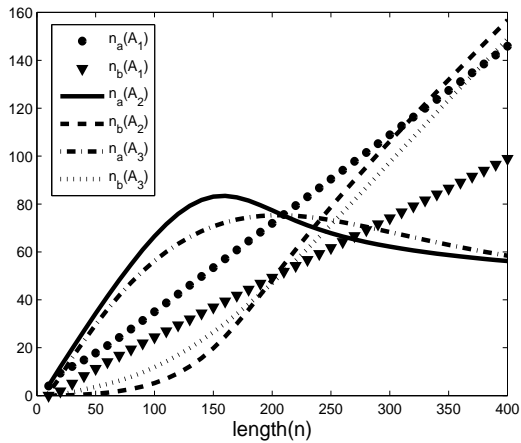}\\
\caption{\label{model2}~Comparison of length dependence under three
different assumptions. Case $A_1$ with $s=1.15$, $\sigma=0.037$,
$w=1.03$, $p=0.072$ and $\eta=1$. Case $A_2$ with $s=1.14$,
$\sigma=0.04$, $w=1.07$, $p=0.0062$ and $\eta=1$. Case $A_3$ with
$s=1.14$, $\sigma=0.0047$, $w=1.03$, $p=0.007$ and $\eta=1$. The
values of parameters in case $A_1$ are the same as in Fig.
\ref{length}. The values of parameters in case $A_2$ and $A_3$ are
chosen to best fit the PDB data between $n=10-200$ .}
\end{figure}

In this paper, we prefer the factor $C_{l_a+l_b}^{l_b}$. Since at
room temperature, $\beta$-strands can not joint freely; and the
connecting patterns are quite limited. Moreover, if we compare case
$A_i(i=1,2,3)$ with PDB data, we can see case $A_1$ gives more
reasonable result(see Fig. \ref{length}). For case $A_2$ and $A_3$,
the number of helical residues $n_a$ decreases with the increment of
length $n$, when the chain is long enough (Fig. \ref{model2}). This
is obviously contradictory to the experimental data.

\end{document}